\documentclass{article}

\usepackage[preprint,nonatbib]{neurips_2021}

\usepackage[utf8]{inputenc} 
\usepackage[T1]{fontenc}    
\usepackage{hyperref}       
\usepackage{url}            
\usepackage{booktabs}       
\usepackage{amsfonts}       
\usepackage{nicefrac}       
\usepackage{microtype}      
\usepackage{xcolor}         

\usepackage{xurl}
\usepackage{amsmath}
\usepackage{amssymb}
\usepackage{amsthm}
\usepackage{amsfonts}
\usepackage{braket}
\DeclareMathOperator*{\argmin}{arg\,min}
\usepackage{upgreek}
\usepackage{dirtytalk}
\usepackage{subfig}
\usepackage[pdftex]{graphicx}

\title{Quantum Federated Learning with Quantum Data}

\author{
  Mahdi~Chehimi\\
  Department of Electrical and Computer Engineering\\
  Virginia Tech\\
  Blacksburg, VA 24060 \\
  \texttt{mahdic@vt.edu} \\
  \And
  Walid~Saad\\
  Department of Electrical and Computer Engineering\\
  Virginia Tech\\
  Blacksburg, VA 24060 \\
  \texttt{walids@vt.edu} \\

}

\begin{document}

\maketitle

\begin{abstract}
Quantum machine learning (QML) has emerged as a promising field that leans on the developments in quantum computing to explore large complex machine learning problems. Recently, some purely quantum machine learning models were proposed such as quantum convolutional neural networks (QCNN) to perform classification on quantum data. However, all of the existing QML models rely on centralized solutions that cannot scale well for large-scale and distributed quantum networks. Hence, it is apropos to consider more practical quantum federated learning (QFL) solutions tailored towards emerging quantum network architectures. Indeed, developing QFL frameworks for quantum networks is critical given the fragile nature of computing qubits and the difficulty of transferring them. On top of its practical momentousness, QFL allows for distributing quantum learning by leveraging existing wireless communication infrastructure. This paper proposes the first fully quantum federated learning framework that can operate over quantum data and, thus, share the learning of quantum circuit parameters in a decentralized manner. First, given the lack of existing quantum federated datasets in the literature, the proposed framework begins by generating the first quantum federated dataset, with a hierarchical data format, for distributed quantum networks. Then, clients sharing QCNN models are fed with the quantum data to perform a classification task. Subsequently, the server aggregates the learnable quantum circuit parameters from clients and performs federated averaging. Extensive experiments are conducted to evaluate and validate the effectiveness of the proposed QFL solution. This work is the first to combine Google’s TensorFlow Federated and TensorFlow Quantum in a practical implementation.
\end{abstract}
\section{Introduction}
\label{introduction}
Recent advances in quantum computing have revolutionized the way computations are done and resulted in quantum computers that solve complex large-scale problems in an extremely faster manner compared to classical computers \cite{nielsen_book}. For instance, in 2019, Google claimed achieving quantum supremacy \cite{quantum_supremacy}, as their quantum computers were able to solve a problem, that would require 10,000 years of computations on a classical computer, within about 200 seconds using a quantum computer. Moreover, IBM is actively developing larger quantum computers \cite{ibm1,ibm_quantum_computer1} with a clear roadmap for scaling up quantum technologies \cite{ibm_quantum_computer1}. These major advances in quantum computing are rapidly leading to the deployment of quantum computers in a wide range of applications including finance \cite{QC_finance}, communication networks, artificial intelligence (AI), and machine learning (ML) \cite{QC_NISQ_era}. 
\subsection{Quantum Machine Learning}
The ability of quantum computers to handle exponential growth in the dimensions of data and perform linear algebra faster and more efficiently than classical computers lead to the blossoming of the field of \emph{quantum machine learning} (QML) \cite{tensorflow_quantum}. For instance, various hybrid quantum-classical ML algorithms have emerged recently, and, when run on quantum computers, achieved superior performance over their corresponding purely classical counterparts. These hybrid algorithms covered different areas in ML such as supervised learning \cite{supervised_QML_book}, quantum support vector machines \cite{quantum_SVM},  unsupervised learning such as clustering \cite{quantum_k_means,NIPS2019_q_means}, and quantum recommender systems \cite{quantum_recommender}. 

With the recent developments of advanced quantum hardware and simulation systems, problems concerned with quantum many-body systems \cite{quantum_many_body_systems} become of particular importance. However, these problems have an extremely complex nature and exponentially large Hilbert spaces which requires exponentially difficult quantum state tomography to translate them into a classical framework efficiently. Moreover, the theoretical analysis of such complex, intrinsically quantum problems is often intractable. All these complex features impose major challenges (see \cite{Cong2019_QCNN}) on classical ML and hybrid QML models, rendering them inefficient when addressing such challenging problems.

These challenges motivated the recent development of a number of \emph{purely quantum ML} models that can, in contrast to non-quantum or hybrid schemes, handle the complex nature of quantum many-body systems. In particular, parametrized quantum circuits (PQC) \cite{PQC_as_ML_models} were proposed as the quantum version of classical neural networks (NNs), forming what is known as \emph{quantum neural networks} (QNN). QNNs are quantum circuits with tunable parameters that can be \say{learned} in a similar fashion to classical NNs.

Several QNN architectures were developed in \cite{farhi2018classification_QNN,Beer2020_QNN} with the goal of performing classification on near-term quantum computers and training QNNs. Moreover, \emph{quantum convolutional neural networks} (QCNNs) were proposed in \cite{Cong2019_QCNN} and achieved a promising performance on classification tasks. Very recently, the authors in \cite{abbas2020power} demonstrated, using actual quantum computers, the advantages of QNNs compared with classical NNs. They verified that, if designed effectively, QNNs achieve a higher effective dimension from an information geometry point of view, in addition to faster training capabilities compared to classical NNs. These findings highly motivate further investigation in the area of QNN given the great potential in this field.

\subsection{Quantum Communication Networks}
Along with the advances in QML, integrating quantum computers in future communication systems is necessary to handle the challenges caused by the rapidly growing volume of data. Moreover, the emerging networks of quantum sensors \cite{guo2020distributed_sensing} and the envisioned quantum internet \cite{quantum_internet2,quantum_internet1} highly motivate designing large-scale distributed quantum networks. Such networks would allow for distributed quantum computing, large-scale quantum communications, and collaborations between quantum clients to perform common tasks. 

In this context, existing quantum communication (QC) networks are particularly suitable for deploying secure communications using the so-called quantum key distribution (QKD) and its variants \cite{qkd}. However, such networks usually rely on photonic quantum hardware that cannot perform strong QML computations, and, thus, they are only suitable for QKD encryption. In fact, the type of quantum computers needed to perform strong computations and run QML frameworks rely on advanced hardware and different technologies such as trapped-ions \cite{trapped_ions1}, quantum dots \cite{quantum_dots1}, superconducting qubits \cite{devoret2004superconducting}, among others. Such technologies require special conditions such as extremely low temperatures, and vibration-free environments to store qubits and effectively maintain their quantum states \cite{nielsen_book}. Thus, \emph{it is much more difficult to transfer the quantum data of QML models using QC channels in an efficient manner}. This is particularly true given the inherent problem of decoherence whereby the quantum data sent in qubits decays gradually as it interacts with the environment, which shortens the lifetime of qubits \cite{nielsen_book}.


Thus, there is a natural need for distributed learning solutions such as federated learning (FL) \cite{kairouz2019advances_FL} in those emerging quantum networks. However, performing a distributed exchange of qubits is not effective for collaborative learning of purely quantum data since it requires the development of extremely complex advanced hardware which will not be available before a few decades \cite{decades}. Given these inherent challenges of distributed quantum systems, one natural research question emerges: \emph{Can we still perform collaborative quantum learning over quantum computing clients while leveraging existing wireless technologies like 5G instead of relying on quantum channels?}

Towards answering this question, we observe that one can exploit the fact that FL algorithms allow the exchange of model parameters, rather than the quantum data itself. As such, it is conceivable that FL can be used to operate over quantum data and, then, leverage classical wireless channels to exchange the model parameters, without the need for a full QC network that is not available. However, remarkably, to the best of our knowledge, no prior work has proposed a thorough, comprehensive framework for implementing FL over purely quantum communication networks, as will be evident from Section \ref{related_works}. As such, this is the key problem we address here.
\subsection{Problem Statement}
\label{problem_statement}
As already noted, emerging large-scale QC networks will require collaborative learning between quantum clients. The qubits used for communication are fragile and susceptible for various losses which renders the efficient transmission of purely quantum data for QML applications a challenging task. Moreover, all of the existing QML models rely on centralized solutions that cannot scale well for future, large-scale and distributed quantum networks. Thus, there is a need for a comprehensive \emph{quantum federated learning} (QFL) framework that allows for distributed quantum learning between purely quantum clients sharing a QML model without the need to transfer the quantum data itself. Such a QFL framework should allow integrating quantum clients in existing classical wireless communication systems. Developing such a QFL framework requires answering many challenging questions:
\begin{itemize}
    \item How can we generate quantum federated datasets, which are non-existent in prior art, in order to practically implement QFL?
    
    \item Are existing classical FL algorithms capable of serializing and learning the quantum circuit parameters, or there is a need for new algorithms?
    
    \item What are the practical challenges and constraints imposed by today's state-of-the-art quantum hardware on applying the proposed QFL framework?    
    
    \item Can the proposed QFL framework handle quantum data with different underlying distributions (e.g., IID vs. non-IID) for different clients?

\end{itemize}
\subsection{Contributions}
The main contribution of this paper is to address the aforementioned challenges by developing a novel comprehensive federated learning framework that allows for distributing quantum learning collaboratively between clients with purely quantum data (e.g., quantum sensors) while leveraging existing classical wireless communication infrastructure. The effectiveness of the proposed approach is validated with extensive experiments and rigorous results that insightfully answer the fundamental challenges posed in Section \ref{problem_statement}. The contributions of this work can be summarized as follows:
\begin{itemize}
    \item We propose the first purely quantum QFL framework for clients with quantum computing capabilities that employ QCNN models to perform a classification task by decentralizing the learning of quantum circuit parameters and performing averaging on the server side.
    
    \item The proposed QFL framework allows for integrating quantum computing clients with the state-of-the-art infrastructure in classical communication networks without relying on quantum channels. Thus, the proposed framework is amenable to practical implementations in existing communication networks.
    
    \item We generate a novel quantum federated dataset which can be used for distributed learning in QC networks. The generated dataset has a hierarchical data format and is the first of its kind in the literature. This dataset can serve as a baseline for future QFL implementations with quantum sensors.
    
    
    \item We conduct the first extensive experiments that combine Google's TensorFlow Quantum (TFQ) \cite{tensorflow_quantum} and TensorFlow Federated (TFF) \cite{TFF}. The developed experiments verify the applicability and effectiveness of the proposed approach. The results show that the proposed QFL framework achieves comparable, and in some cases superior, performance compared to the centralized, purely QML setup. Our results also show that the model can handle IID and non-IID quantum data. A key finding is that classical FL algorithms can be applied to decentralize the learning in purely quantum QML applications.

\end{itemize}
\subsection{Related Works}
\label{related_works}
A handful of prior works \cite{distributed_secure_QML,narottama2020quantum,yang2020decentralizing_VFL_speech,chen2021federated} exist on decentralized QML models, but those works neither address the previously posed challenges nor provide a solution to practically distribute quantum learning algorithms that are tailored towards emerging large-scale purely QC networks with purely quantum QML. First, the authors in \cite{distributed_secure_QML} proposed a protocol using which a single client that does not have enough quantum capabilities performs a classical ML task (vector classification) by communicating with a quantum server to run a small-scale QML model and send the learned parameters back to the client. The main objective of \cite{distributed_secure_QML} is to preserve privacy of the communication link while training the classical ML model. However, this work considers a single client, and it does not scale-up to emerging quantum networks that encompass multiple distributed clients as we consider here. Also, this prior art \cite{distributed_secure_QML} does not consider the deployment of quantum computing clients nor the  integration of quantum data-based QML networks which is a challenging task that we will address.

Meanwhile, a concise vision of a QFL architecture for general optimization in wireless communication networks is discussed in \cite{narottama2020quantum}. This prior work considers a radically different scenario than the one treated here. In particular, it considers a wireless network of classical non-quantum mobile users, communicating with access points that run a shallow QNN model for optimizing the wireless network. The access points use FL and share the learning parameters with a server having another deeper QNN model. However, the work in \cite{narottama2020quantum} only discusses a conceptual architecture without any implementation, verification, or concrete results. Moreover, it does not consider quantum clients or quantum data in the communication network, and it relies solely on classical data. 

Meanwhile, the authors in \cite{yang2020decentralizing_VFL_speech} considered a vertical federated learning approach for decentralized feature extraction in automatic speech recognition tasks. Their model includes a quantum server that uses a QCNN for feature extraction. However, this approach is totally different from our proposed framework as it primarily studies the usage of QCNNs for extracting useful features from classical non-quantum data, and only assumes the server to have quantum capabilities in the communication network.

Finally, the most relevant prior work is the work in \cite{chen2021federated} which considered a hybrid quantum-classical ML model trained in a federated setup. The work in \cite{chen2021federated} is different from our proposed QFL framework since it adopts a transfer learning approach where a pre-trained CNN model is used to extract features from classical data and compress it into a vector passed to a QNN to make predictions. Although the authors in \cite{chen2021federated} discuss federating the QML models, their implementation uses classical data and has a very limited contribution to the purely quantum setup since it does not address the biggest challenges such as the lack of a purely quantum federated dataset in the literature and the lack of a holistic implementation.

Clearly, there is no prior work that addresses QFL while leveraging purely quantum data learning. This gap in the literature must be extensively addressed as it could provide a breakthrough in the way QC networks are looked at. 

The rest of this paper is organized as follows. Section \ref{Problem_setup} describes the problem setup and the adopted QFL model in details. Next, the proposed quantum federated dataset is developed and the proposed process to generate it is presented in Section \ref{data_section}. In Section \ref{experiments}, we conduct the experiments and discuss the key results. Conclusions are drawn in Section \ref{conclusion}. Finally, insights on the challenges facing the proposed QFL framework and its broader impact are presented in Section \ref{future}.
\section{Problem Setup}
\label{Problem_setup}
\begin{figure}[t]
  \begin{center}
    \includegraphics[scale=0.35]{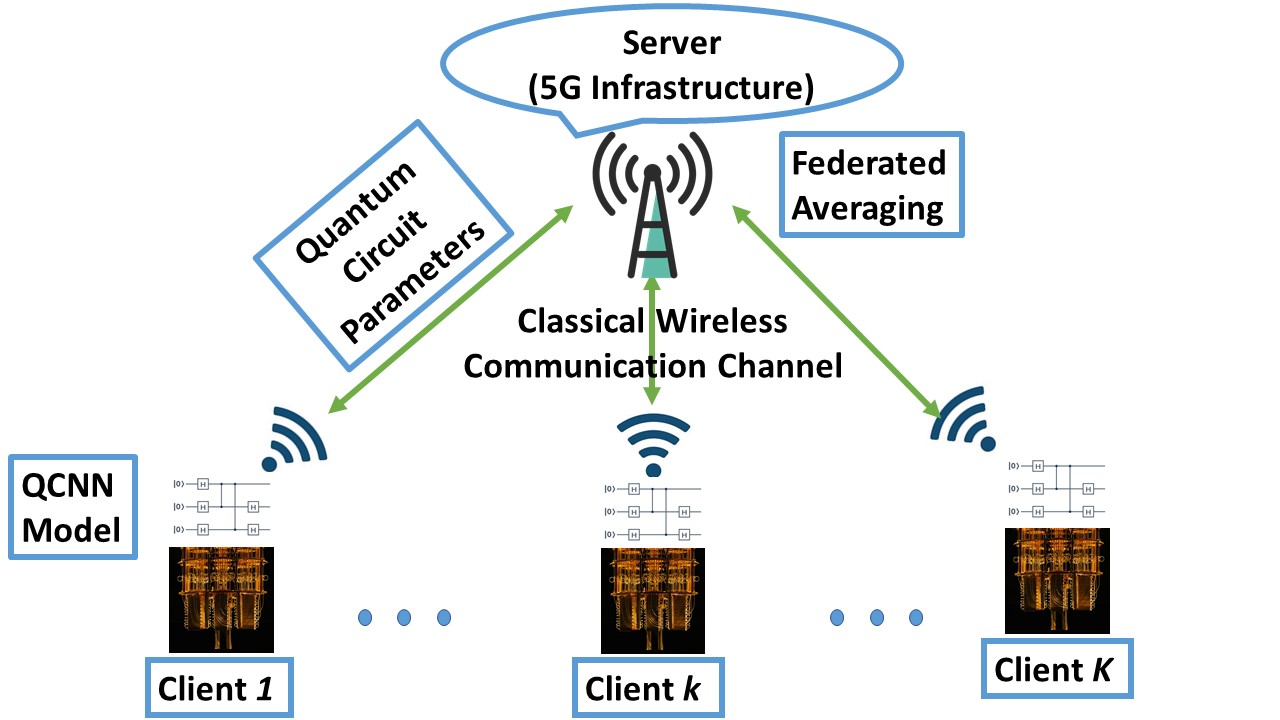}
    \caption{Problem Setup.}\label{fig_problem_setup}
  \end{center}
\end{figure}

We consider a quantum network setup (Figure \ref{fig_problem_setup}) consisting of a server and a set $\mathcal{K}$ of $K$ quantum computing clients sharing a QCNN model such as the one proposed in \cite{Cong2019_QCNN}. Each client generates its own quantum data locally and trains a local QCNN model to perform binary classification. The generated data consists of excitations for an $N$-qubits quantum cluster state fed as labeled inputs in pairs $(\ket{\psi_m},y_m)$: $m = 1,2,...,M$, where $\ket{\psi_m}$ represents the $m$-th sample input quantum state, $y_m$ is a binary label that classifies whether the cluster state is excited (takes value of 0) or not (takes value of 1), and $M$ is the number of data samples. Such a setup is typical for quantum sensor networks and will be further analyzed in Section \ref{data_section}.

In an analogous manner to classical convolutional neural networks, and following a translationally invariant behavior, the QCNN model includes a sequence of quantum convolution layers followed by quantum pooling layers. After incorporating sufficient layers, a quantum fully connected layer is applied, and predictions are made by performing quantum measurement of qubits. 

More formally, each convolution layer $C \in \mathcal{L}_c$, with $\mathcal{L}_c$ being the set of convolution layers, is a single quasi-local unitary. This could represent any quantum gate, since unitarity is the main constraint that must be imposed on any matrix that is used to represent a quantum gate \cite{nielsen_book}. In this regard, a given matrix $\boldsymbol{U}$ corresponding to a single qubit gate is said to be \emph{unitary} if $\boldsymbol{U}^\dagger \boldsymbol{U} = \boldsymbol{I}$, where $\boldsymbol{U}^\dagger$ is the adjoint of $\boldsymbol{U}$, and $\boldsymbol{I}$ is the identity matrix. A unitary is called quasi-local if a quasi-local Hamiltonian generated that unitary as explained in \cite{quasi_local_unitary}. 

In a quantum pooling layer $P \in \mathcal{L}_p$, with $\mathcal{L}_p$ being the set of pooling layers, the system size (and the degrees of freedom) are reduced, resulting in non-linearities. This is done by measuring some qubits and using the measurements to determine unitary rotations and apply them to other close qubits. After obtaining a system size that is sufficiently small, a quantum fully connected layer $F$ is applied on the remaining qubits and is represented by a unitary. Finally, measurement is performed on a specific number of qubits at the output to produce the QCNN's prediction.

At each client, the learnable parameters are the entries of all unitaries ,i.e., the quantum circuit parameters which are classical values. Let $\text{\boldmath$\theta$}^k = (C,P,F)$ where $k\in\mathcal{K}$ be the vector of all QCNN's model parameters for client $k$, then the predicted output value by the QCNN model $f$ to input quantum state $\ket{\psi_m}$ for client $k$ is denoted by $f_{\text{\boldmath$\theta$}^k}(\ket{\psi_m})$. Initially, the model parameters are initialized, and, then, they are updated in order to minimize the following mean squared error (MSE) loss function:
\begin{equation}
\argmin_{\boldsymbol{\theta}^k} \mathcal{J}(\boldsymbol{\theta}^k) := \frac{1}{2M} \sum_{m=1}^{M}{(y_m - f_{\boldsymbol{\theta}^k}(\ket{\psi_m}))^2}.
\end{equation} 
In order to benefit from each others experience and data, all $K$ clients will collaborate in training the same QCNN model. Such a collaboration can be beneficial for emerging applications such as large-scale quantum communication networks. To perform this collaboration, we use the proposed QFL framework. In this context, the general setup of QFL follows a similar structure to classical FL \cite{yu2020federated_icml}. The collaborative learning is done by using existing wireless communication technologies such as the 5G cellular infrastructure to send the clients' \say{classical} model parameters to the server. This setup will, in turn, allow the system to perform decentralized quantum data learning while using existing classical wireless links, a task that is extremely difficult to achieve efficiently if the clients needed to send the quantum data itself to the server. 

Each round $h$ of the QFL training starts with the server sending its current version of the model parameters $\text{\boldmath$\theta$}_h^s$ to all $K$ clients. Each client $k \in \mathcal{K}$ uses its local quantum data to run an optimization algorithm such as stochastic gradient descent (SGD), in order to update its model parameters according to the following equation:
\begin{equation}
    \boldsymbol{\theta}_h^k = \boldsymbol{\theta}_h^s - \eta\cdot\nabla_{\boldsymbol{\theta}_h^k} \mathcal{J}(\boldsymbol{\theta}_h^k), 
\end{equation}
for every $k\in \mathcal{K}$ and  with $\eta$ being the learning rate.
Next, each client $k$ sends its updated model parameters back to the server which aggregates the parameters from all clients. Then, the server applies the popular Federated Averaging\footnote{Other advanced FL algorithms can also be accomodated as part of our framework.} FL algorithm \cite{FedAvg} to estimate an average update of the model parameters and send the newly updated parameters to all clients according to the following rule:
\begin{equation}
    \boldsymbol{\theta_{h+1}^s} = \sum_{k\in \mathcal{K}}{w_k\cdot\boldsymbol{\theta_h^k}},   
\end{equation}
where the weighting vector $\boldsymbol{w} = (w_1, w_2,...,w_k)$ is assigned by the server to weigh different clients in the network. This process is repeated until convergence. 

Given this setup, our key goal is to implement the QFL framework over a quantum network in which quantum learning is performed in a decentralized manner using classical wireless communication infrastructure. To do so, we next generate a novel quantum federated dataset, and, then, we implement the QFL framework. This will constitute the first implementation of such a system that combines Google's TFQ and TFF.
\section{Quantum Federated Dataset}
\label{data_section}
Given the lack of existing quantum federated dataset in literature, our proposed QFL framework must begin by generating the first quantum federated dataset that can be used for distributed learning in quantum networks. The generated dataset has a hierarchical data format and consists of purely quantum data. Next, we describe the proposed steps for generating the dataset.
\subsection{Quantum Data}
A variety of purely quantum data exists for different quantum many-body systems \cite{quantum_many_body_systems}. The data is typically obtained using different quantum devices or complex simulations of quantum systems.\footnote{Note that classical data can be encoded into quantum states and fed to quantum computers as quantum data as discussed in \cite{effect_data_encoding_variational_QML} This classical data encoding task is done by performing quantum state preparation functions or by utilizing quantum feature maps.} For instance, \cite{Cong2019_QCNN} considered symmetry-protected topological phases \cite{Gu_2009_SPT1,Schuch_2011_SPT2} as input data to be classified by a QCNN model. We adopt the simpler, yet practical, form of quantum data consisting of quantum clusters inspired from \cite{TFQ_qcnn}.  

The proposed quantum dataset consists of excitations of quantum cluster states \cite{Nielsen_2006_cluster_states_quantum_computation,Briegel2009_cluster_states}, labeled as either excited or not based on the rotations of the qubits. This type of data is of important use for quantum sensing networks like the ones applied for metrology \cite{Friis_2017_cluster_states_metrology}. In addition, quantum cluster states are particularly important for our tackled practical problem of QC networks. This is due to their applicability in distributed quantum computing networks, and their ability to teleport quantum states between quantum clients communicating through a quantum channel \cite{cluster_states_communications}. Thus, the adopted data type perfectly fits our targets and allows for future extensions to quantum networks incorporating both classical and quantum clients.
\subsection{Federated Dataset Generation}
\subsubsection{Generating Single Client Data}
We used TFQ and Google's framework for quantum circuit programming: Cirq \cite{cirq} for generating the quantum data of each client. We begin by generating a rectangular grid of $1\times8$ qubits using Cirq, such a size is reasonable for QML simulations and is sufficient for validating the proposed QFL framework. Then, we create cluster states as TFQ circuits consisting of the Hadamard and Controlled-Z quantum gates \cite{nielsen_book}, and apply the circuit on the generated qubits. In order to define the excitations of cluster states, we observe that most quantum gates operating on a single qubit can be described as rotations around an axis in the Bloch Sphere \cite{lu2005bloch1,glendinning2005bloch2}. Thus, they are usually referred to by their axis of rotation \cite{cirq}. As proposed by \cite{TFQ_qcnn}, we consider excitations to be represented by rotation gates around the x-axis of the Bloch Sphere ($R_x$ quantum gates \cite{nielsen_book}). An excitation of the cluster state is declared if a large enough rotation is achieved and the state is labeled with 1. In case the rotation is not sufficiently large, the state is declared as unexcited, and is labeled with a 0.
\subsubsection{Generating Federated Data}
The previously described quantum data has input represented by quantum circuits. In order to be able to store the data, the quantum circuit is transformed into a tensor that is represented by strings. In fact, these strings represent an encoding of the serialized binary data of the quantum circuits with TensorFlow data type \say{lS5000}. We particularly generate a hierarchical data format version 5 (HDF5) federated dataset file which includes examples of 30 clients (Any number of clients can be considered, see Section \ref{experiments}). Each client has its own quantum dataset consisting of 160 serialized binary data input vector of a single feature, and a labels vector of integers of 0s and 1s. 

Moreover, we generate a quantum federated dataset consisting of non-IID clients' datasets. Then, we conduct key experiments on this setup to compare the performance with the IID case originally followed. The obtained results are discussed in the next section.

\section{Experiments} 
\label{experiments}
In this section, we present the experimental results for the proposed QFL framework. First, we provide thorough details of the experimental setup. Then, we conduct extensive experiments to verify the applicability and effectiveness of our proposed framework. 



\subsection{Experimental Setup}\label{sub_exp_setup}
\textbf{Implementation}. We use the TFQ and TFF frameworks to implement our proposed QFL framework, and we build upon the implementations in \cite{TFQ_qcnn,FLW}. We run our simulations on Google's Cloud Platform known as \say{Google Colaboratory \cite{collab}} using CPU computing nodes. The implementation begins with generating the quantum federated dataset described in Section \ref{data_section}. 

Compared to classical FL scenarios, it is natural to assume that the number of quantum computing clients in a quantum network will be in the range of tens of clients. It is typically assumed that, at a given point in time, only a subset of clients is available for training. However, for the ease of simulations and since the number of clients is small, we assume that all quantum clients available for training, and we reserve the data of a small portion of clients for testing. Since TFF is currently only available in simulation environments, then each client's data is assumed to be available locally.

As is typical in classical CNN models, the adopted QCNN architecture consists of quantum convolution layers, each followed by a quantum pooling layer. Then, a quantum fully connected layer is included, and finally, the measurement is performed on the last layer. The width of the QCNN is not an optimization parameter since it solely relies on the number of qubits in the system, due to the reduction in size that quantum pooling layers introduce to the input qubits. In our case, and since we consider 8 qubits in the QFL setup (this is a typical value for quantum sensing networks), we found that having three pairs of quantum convolution-pooling layers, with 64 learnable parameters, is the most suitable QCNN architecture that fits our setup.

\textbf{Optimizers and hyperparameters}. We compare the performance of the QFL framework under various client optimizers (Adam, SGD, and RMSprop) while fixing the server's optimizer to SGD with a learning parameter equals to 1, since it is only performing averaging. The learning rates of the clients are varied, and their impact on the performance is discussed in Section \ref{sec_optimizers}.

\textbf{Validation metrics}. While it is typical in centralized ML models to use validation data when training, this sort of data is inaccessible in the QFL setup. Thus, a subset of clients will be specified for testing and validating the performance of the trained QFL framework. The validation metric used for the testing clients is the binary accuracy metric with a threshold of 0.5, which is typical for binary classification problems.

\subsection{Results and Discussion}
\subsubsection{Impact of Number of Clients}
\begin{figure}[t]
  \begin{center}
    \includegraphics[scale=0.33]{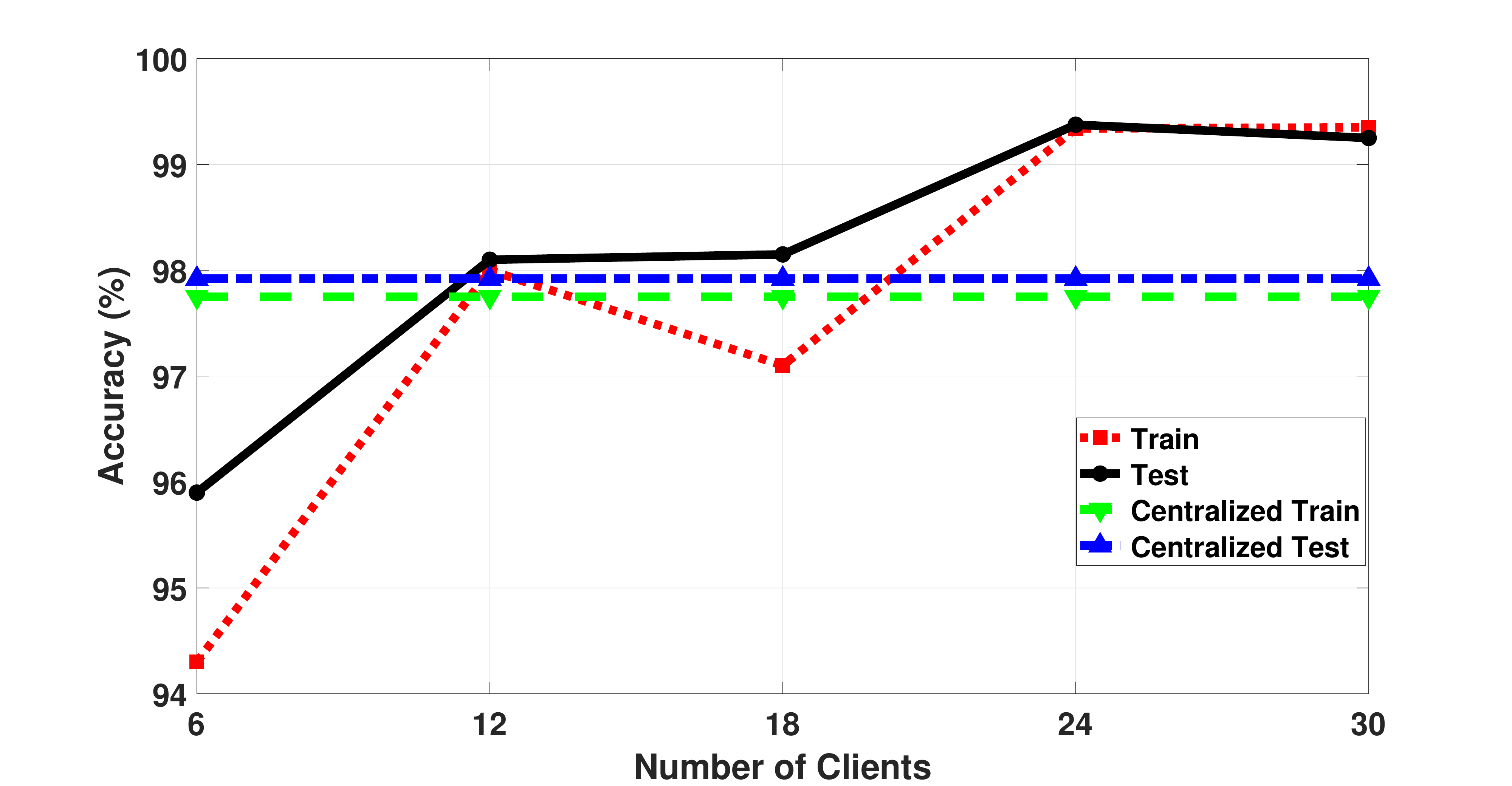}
    \caption{Evaluation of QFL accuracy as the number of clients varies.}\label{fig_num_clients}
  \end{center}
\end{figure}

We begin our experiments by analyzing the impact of the number of available clients in the QFL network. In Figure \ref{fig_num_clients}, we compare the achieved testing accuracy when the quantum federated dataset has 1 (centralized), 6 (4 for training, 2 for testing), 12 (9 for training, 3 for testing), 18 (14 for training, 4 for testing), 24 (19 for training, 5 for testing), and 30 (25 for training, 5 for testing) clients while fixing the number of data samples available to each client to 160 samples. We observe that, in general, as the number of participating clients in the QFL setup increases, a higher testing accuracy is achieved without overfitting the training data. The reason why the case of 6 clients achieves a smaller accuracy is because the number of clients in a federated setup must be large enough in order to achieve efficient learning.


\subsubsection{Impact of the Size of Clients' Data Samples}
\begin{figure}[ht]
  \begin{center}
    \includegraphics[scale=0.33]{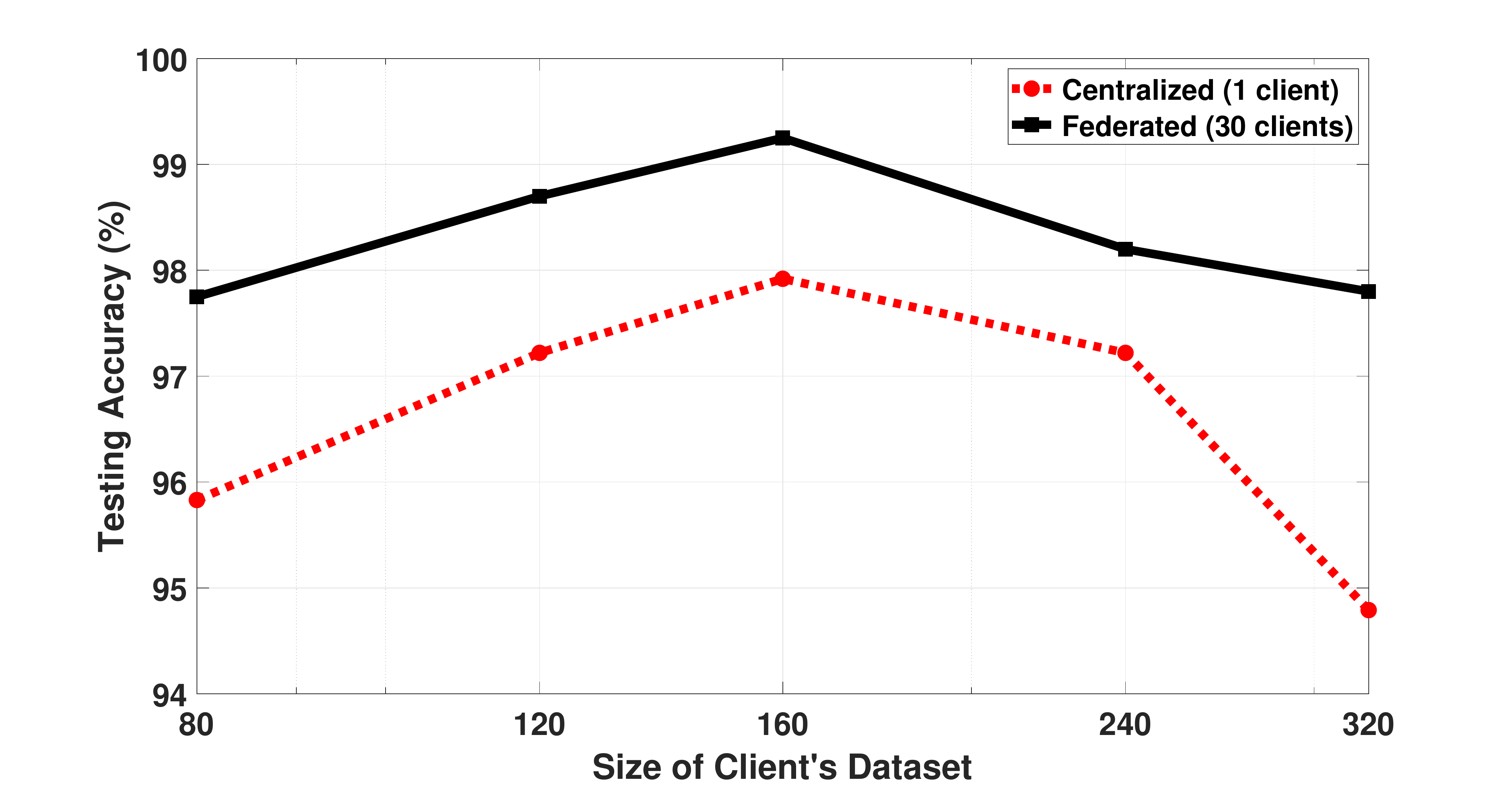}
    \caption{Size of client's dataset vs testing accuracy.}\label{fig_data_size}
  \end{center}
\end{figure}

In Figure \ref{fig_data_size}, we consider a QFL network of 30 active clients and compare the achieved testing accuracy with the centralized case (single client) while varying the size of the individual client's datasets. Since the adopted QCNN is shallow and the number of trainable parameters is small (64), we observe that increasing the size of the dataset does not necessarily guarantee an improvement in the performance. In fact, as long as each client has enough data, increasing the size of the dataset may slightly increase or decrease the achieved testing accuracy. Another important observation is that, in this network, the federated framework achieves a superior performance compared to the centralized case. This is because each client in the federated setup benefits from the data of the other clients in the learning process, which enhances the performance. 

\subsubsection{Impact of Optimizers and Learning Rates}\label{sec_optimizers}
\begin{figure}[ht]
  \begin{center}
    \includegraphics[scale=0.33]{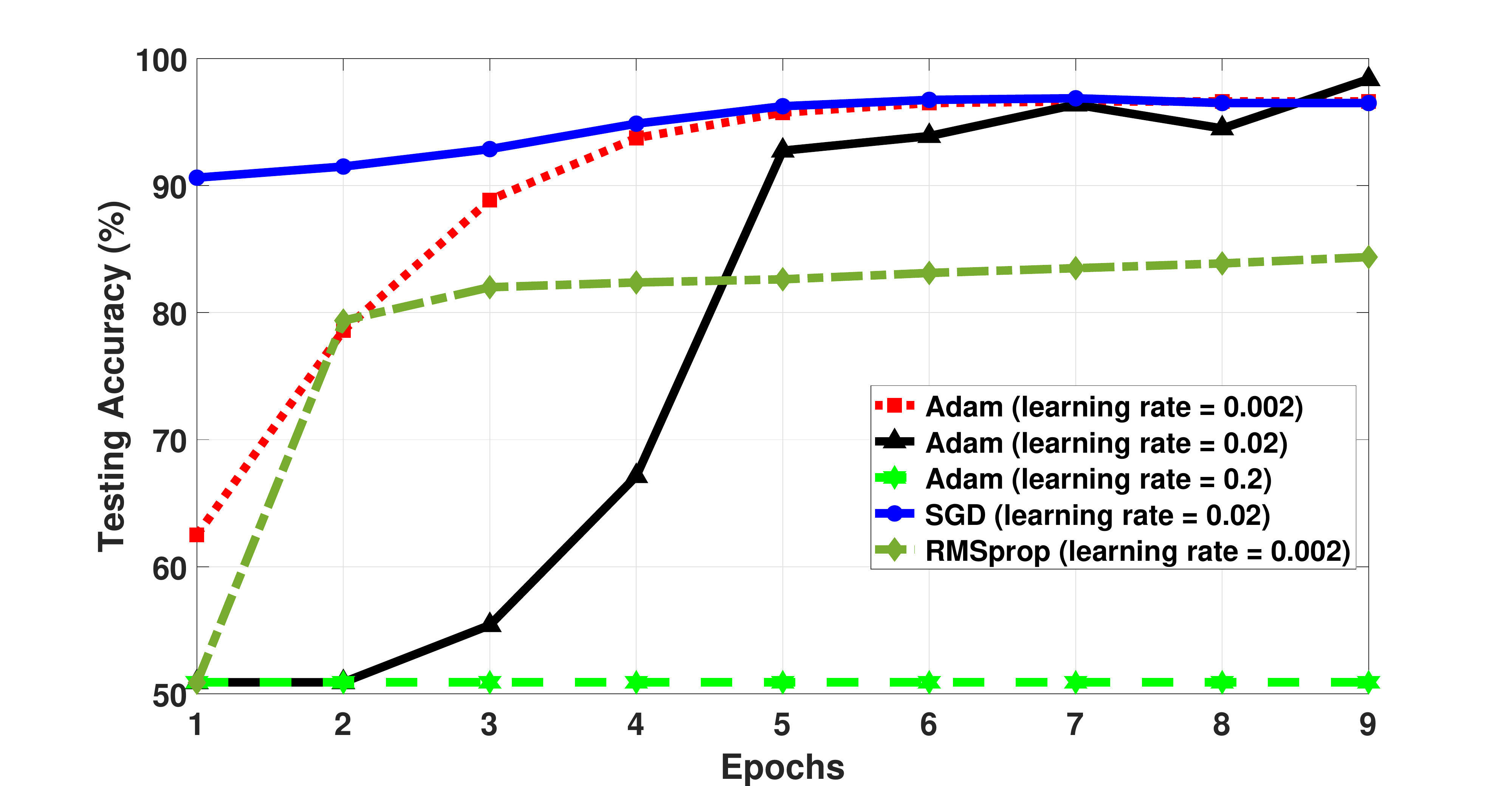}
    \caption{Evolution of the testing accuracy of different optimizers over the training epochs.}\label{fig_different_optimizers}
  \end{center}
  \vspace{-0.15in}
\end{figure}

In Figure \ref{fig_different_optimizers}, we compare the performance of the QFL framework with different optimizers and learning rates. We observe that the RMSprop optimizer with a learning rate of $0.002$ is slow compared to the other optimizers and converges at a smaller testing accuracy. For the SGD optimizer with a learning rate of $0.02$, we observe that it is the only one that achieves a high accuracy from the first epoch, and it converges to a value around $96.5\%$. However, the Adam optimizer with a learning rate of $0.02$ converges to a higher testing accuracy after few training epochs. Finally, we observe that a learning rate of $0.2$ for Adam optimizer is very large that it cannot learn, while a learning rate of $0.002$ results in a smoother curve at the expense of a smaller accuracy.

\subsubsection{Impact of Non-IID Data}
When generating the quantum cluster states from the input qubits, the rotation values fed to the $R_x$ gate were drawn from a uniform distribution between $-\pi$ and $\pi$. In order to vary the underlying distribution of the clients' data, we consider generating the data for half of the clients using a truncated normal distribution, so that we generate non-IID quantum data. After generating the new dataset, we compare the performance of the QFL framework on the IID and non-IID federated datasets for a network of 30 clients (25 for training, 5 for testing) with 160 data samples each. In Table \ref{non_iid}, we show the testing accuracy and MSE loss for both datasets and observe that the proposed QFL framework achieves a similar performance on both IID and non-IID quantum federated datasets.

\begin{table}
  \caption{Performance comparison between IID and non-IID data}
  \label{non_iid}
  \centering
  \begin{tabular}{lll}
    \toprule
    Dataset     & Testing Accuracy     & Testing MSE \\
    \midrule
    IID Data & 99.25  & 5.66    \\
    Non-IID Data & 98.625 & 6.57     \\
    \bottomrule
  \end{tabular}
\end{table}

\subsubsection{Error Bars}\label{subsec_error_bars}
\begin{figure}[ht]
  \begin{center}
    \includegraphics[scale=0.33]{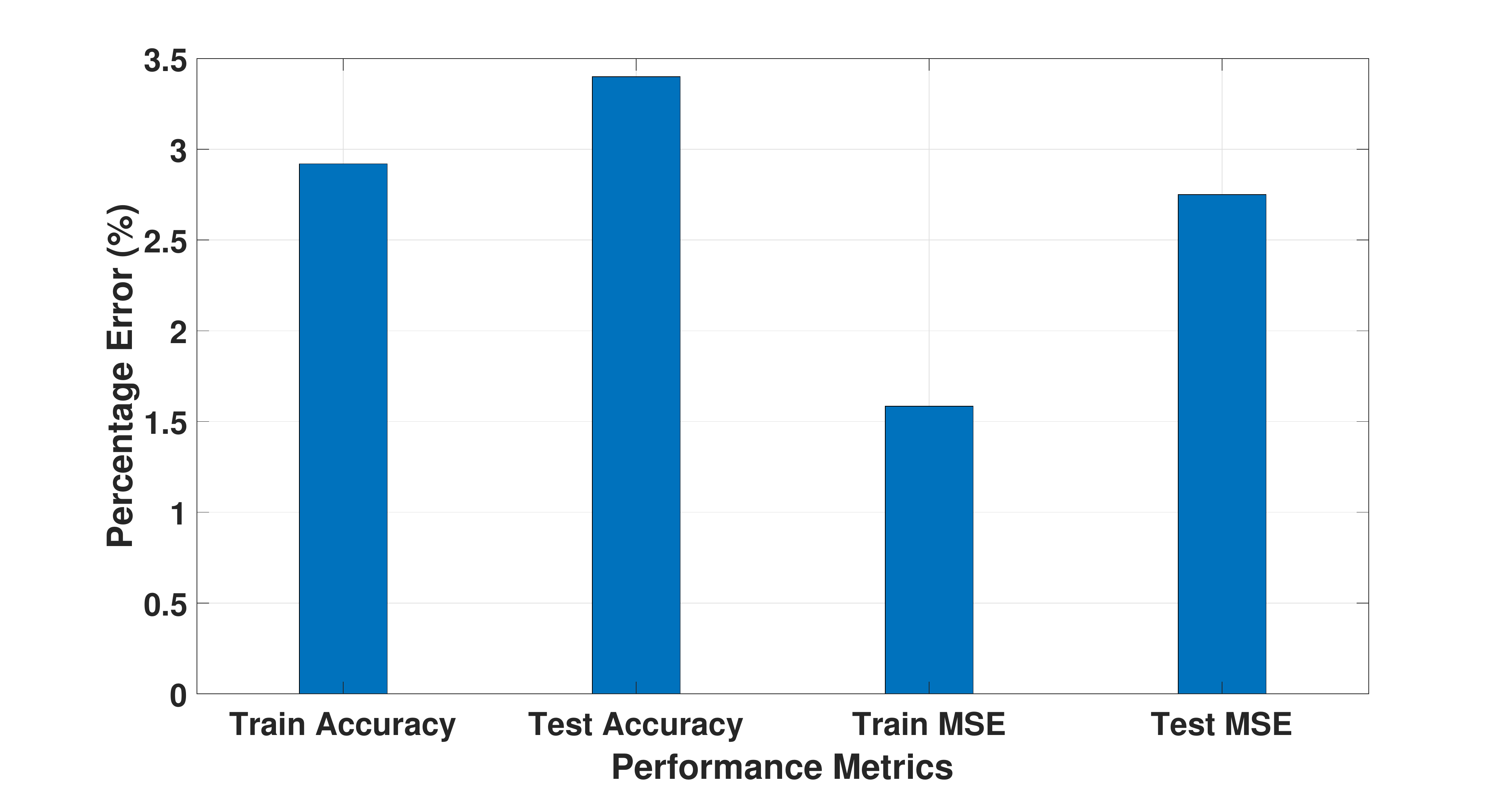}
    \caption{Error bars from the random seed after running experiments multiple times.}\label{fig_error}
  \end{center}
\end{figure}

In Figure \ref{fig_error}, we show the error bars for the different performance metrics for both training and testing cases. The tested network consists of 30 clients each with a dataset of 160 samples and the data of 5 clients is used for testing. Figure \ref{fig_error} shows the errors caused by the random seed after running multiple experiments for the same network, and we can see that the errors are below $5\%$ which is acceptable.   

\section{Challenges and Future Outlook}
\label{future}
The state of the art near-term-intermediate-scale quantum (NISQ) hardware only allow for processing a small number of qubits because of the difficulty in performing quantum error correction to minimize the losses. This is a challenge that renders the practical deployment of large-scale purely quantum QML models, like our proposed QFL framework, a difficult task. Also, the variety of technologies and the different quantum capabilities of devices in a quantum network may make the adoption of a unified QML model difficult, and, thus, the practical implementation of our proposed QFL framework challenging. 

However, our proposed QFL framework paves the way for integrating quantum devices with quantum data with existing wireless networks. As a result, we anticipate that this can lead to a blossoming of new applications that are coupled with new research problems in the areas of networking, quantum hardware, and wireless sensing. This includes designing efficient optimization algorithms that account for networks including both quantum and classical computing clients. This is a great breakthrough that allows leveraging the powerful computing capabilities of quantum computers in today's communication networks. Our proposed QFL framework allows for training the QML models inside future 6G communication networks which is very promising. Finally, in order to add a new layer of security to our porposed QFL framework, one can investigate the use of quantum cryptographic schemes to encrypt the classical learning parameters in the QFL setup before sending them to the server, and vice versa. Since the clients have quantum capabilities, integrating QFL with QKD is an inteteresting challenging problem that is worth investigation in the future.

\section{Conclusion}
\label{conclusion}
In this paper, we have proposed a novel framework for quantum federated learning that allows implementing scalable distributed quantum learning over quantum data without the need to send qubits, but by leveraging classical wireless networks. To implement this framework, we have generated the first quantum federated dataset in the literature and performed a unique implementation that combines TensorFlow Quantum and TensorFlow Federated. We have conducted extensive experiments to verify the applicability and effectiveness of our proposed framework. The experimental results validate the effective behavior of the proposed QFL framework using the federated averaging algorithm.

\bibliographystyle{IEEEtran}
\bibliography{References}

\end{document}